\documentclass[twocolumn]{article} 

\long\def\comment#1{}

\makeatletter
\def\@normalsize{\@setsize\normalsize{10pt}\xpt\@xpt
\abovedisplayskip 10pt plus2pt minus5pt\belowdisplayskip
\abovedisplayskip \abovedisplayshortskip \z@
plus3pt\belowdisplayshortskip 6pt plus3pt
minus3pt\let\@listi\@listI}

\def\subsize{\@setsize\subsize{12pt}\xipt\@xipt}
\def\section{\@startsection {section}{1}{\z@}{1.0ex plus
1ex minus .2ex}{.2ex plus .2ex}{\large\bf}}
\def\subsection{\@startsection
   {subsection}{2}{\z@}{.2ex plus 1ex} {.2ex plus .2ex}{\subsize\bf}}
\makeatother

\usepackage{amsfonts, amssymb, amsmath, amscd}
\usepackage{graphicx}
\usepackage[dvips]{epsfig}
\usepackage{color}
\usepackage{subfigure}
\usepackage{stfloats}
\usepackage{cuted}

\newcommand{\param}{\color{black}}
\newcommand{\qoi}{\color{black}}

\graphicspath{{figs/}}

\newcommand{\mbb}[1]{\mathbb{#1}}

\newcommand{\vvvert}{|\kern-1pt|\kern-1pt|}

\begin{document}

\date{}

\title{\huge \bf Application of Predictive Model Selection to Coupled Models}

\author{Gabriel Terejanu\thanks{E-mail address: terejanu@ices.utexas.edu. Address: 
The University of Texas at Austin, 
1 University Station, C0200 
Austin, TX 78712, USA } ~,~
Todd Oliver\thanks{E-mail address: oliver@ices.utexas.edu} ~,~ Chris Simmons\thanks{E-mail address: csim@ices.utexas.edu}\\
Center for Predictive Engineering and Computational Sciences (PECOS) \\
Institute for Computational Engineering and Sciences (ICES) \\
The University of Texas at Austin
}

\maketitle
\thispagestyle{empty}

{\hspace{1pc} {\it{\small Abstract}}{\bf{\small---
A predictive Bayesian model selection approach is presented to discriminate 
coupled models used to predict an unobserved quantity of interest (QoI). 
The need for accurate predictions arises in a variety of
critical applications such as climate, aerospace and defense.
A model problem is introduced to study the prediction yielded by the coupling 
of two physics/sub-components. 
For each single physics domain, a set of model classes and a set of sensor 
observations are available. A goal-oriented algorithm using a predictive approach to 
Bayesian model selection is then used to select the combination of single physics 
models that best predict the QoI. It is shown that the best coupled model for 
prediction is the one that provides the most robust predictive distribution 
for the QoI.

 }}
}


\section{Introduction}
\label{sec:intro}

With the exponential growth of available computing power and the continued development 
of advanced numerical algorithms, computational science has undergone a revolution in 
which computer models are used to simulate increasingly complex phenomena. Additionally, 
such simulations are guiding critical decisions that affect our welfare and security, 
such as climate change, performance of energy and defense systems and the biology of 
diseases. Reliable predictions of such complex physical systems requires sophisticated 
mathematical models of the physical phenomena involved. But also required is a 
systematic, comprehensive treatment of the calibration and validation of the models, 
as well as the quantification of the uncertainties inherent in such models. 
While recently some attention has been paid to the propagation of uncertainty, 
considerably less attention has been paid to the validation of these complex, 
multiphysics models. This becomes particularly challenging when the quantity of 
interest (QoI) cannot be directly measured, and the comparison of model predictions
with real data is not possible. Such QoIs may be the catastrophic
failure of the thermal protection system of a space shuttle reentering the atmosphere
or the different performance characteristics of nuclear weapons in order to maintain
the nuclear stockpile without undergoing underground nuclear testing.

In this paper, we present an intuitive interpretation of the predictive model selection
in the context of Bayesian analysis. While the predictive model selection is not 
an new idea, see Refs.\cite{Geisser1979, Martini1984, Trottini2002}, here we emphasize 
the connection between the QoI-aware evidence and the Bayesian model averaging
used for estimation. This new interpretation
of the Bayesian predictive model selection reveals that the best model for prediction
is the one which provides the most robust predictive probability density function (pdf)
for the QoI. Also, the latest advances in Markov Chain Monte Carlo (MCMC) 
\cite{Cheung_2009C} and estimators based on the $k$-nearest neighbor \cite{Wang2006} 
are used to compute the information theoretic measures required in the problem 
of predictive selection of coupled models. It is further argued that equivalence
between predictive model selection and conventional Bayesian model selection can be
reached by performing optimal experimental design \cite{Terejanu2011_ExpDes} 
for model discrimination.
The structure of the paper is as follows: first the selection problem of coupled models
is stated in Section \ref{sec:prob}. The conventional Bayesian model selection is
described in Section \ref{sec:Bayes} and the extension to QoI-aware evidence is
derived in Section \ref{sec:predictive}. The model problem and numerical results
are presented in Section \ref{sec:model} and Section \ref{sec:results} respectively.
The conclusions and future work are discussed in Section \ref{sec:conclusions}.

\section{Problem Statement}
\label{sec:prob}

\begin{figure}[h]
  \centering
  \includegraphics[height=.8\linewidth]{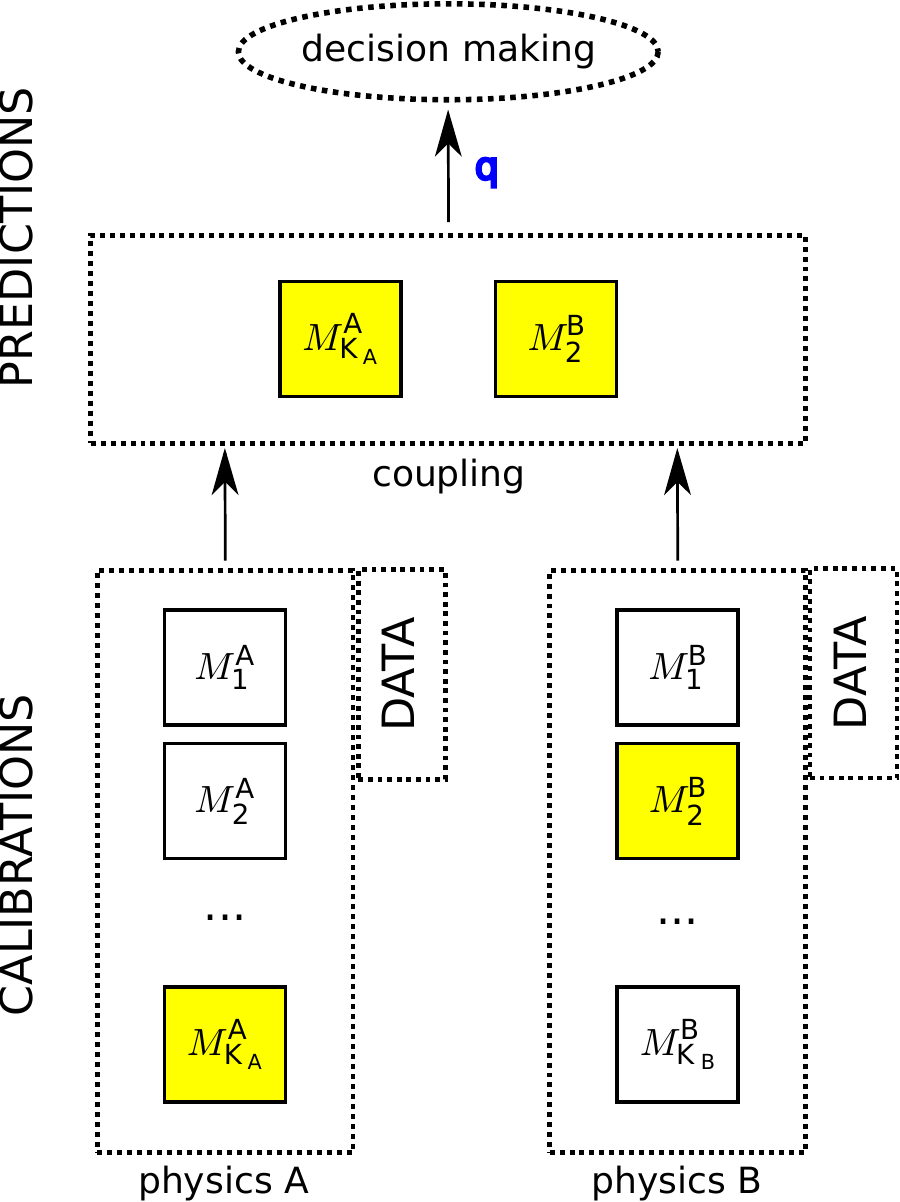}
  \caption{Predictive Selection for Coupled Models}\label{fig:problem}
\end{figure}

Here we are interested in the prediction of a coupled model. The problem of selecting the best coupled model in the context of the QoI is to find the combination of single physics models that best predict an unobserved QoI in some sense, see Fig. \ref{fig:problem}. Thus at the single physics level, we have two physics A and B, each with a model class set, $\mathcal{M}^A$ and $\mathcal{M}^B$, and a set of observations $D^A$ and $D^B$ respectively. The cardinality of the two sets of model classes are $|\mathcal{M}^A|=K_A$, $|\mathcal{M}^B|=K_B$, and the the definition of model classes in each set is given by the state equations and measurement models as follows,
\begin{eqnarray}
  M_i^A : \begin{cases}
    \mathbf{r}_i^A(\mathbf{u}_i^A,{\param{\boldsymbol{\theta}}_i^A}) = \mathbf{0}  \nonumber \\
    \mathbf{y}^A = \mathbf{y}_i^A(\mathbf{u}_i^A,{\param{\boldsymbol{\theta}}_i^A}) \nonumber
  \end{cases}
  M_j^B : \begin{cases}
    \mathbf{r}_j^B(\mathbf{u}_j^B,{\param{\boldsymbol{\theta}}_j^B}) = \mathbf{0} \nonumber \\
    \mathbf{y}^B = \mathbf{y}_j^B(\mathbf{u}_j^B,{\param{\boldsymbol{\theta}}_j^B}) . \nonumber
  \end{cases}
\end{eqnarray}

All the possible couplings of single physics models yield the set of coupled model classes 
$\mathcal{M} = \{ M_{ij} = (M_i^A,M_j^B) | M_i^A \in \mathcal{M}^A, M_j^B \in \mathcal{M}^B \}$ with cardinality $|\mathcal{M}| = K_A K_B$. The definition of a coupled model class in this set is given by the state and measurement equation, and in addition we also have the model for the QoI,
\begin{eqnarray}
  M_{ij} : \begin{cases}
    \mathbf{r}( \mathbf{u}_i^A, \mathbf{u}_j^B, {\param{\boldsymbol{\theta}}_i^A}, {\param{\boldsymbol{\theta}}_j^B} ) = \mathbf{0} \nonumber \\
    \mathbf{y}^{AB} = \mathbf{y}_{ij}^{AB}( \mathbf{u}_i^A, \mathbf{u}_j^B, {\param{\boldsymbol{\theta}}_i^A}, {\param{\boldsymbol{\theta}}_j^B} ) \nonumber \\
    {\qoi\mathbf{q}} = \mathbf{q}_{ij}^{AB}( \mathbf{u}_i^A, \mathbf{u}_j^B, {\param{\boldsymbol{\theta}}_i^A}, {\param{\boldsymbol{\theta}}_j^B} ) . \nonumber
  \end{cases}
\end{eqnarray}

Having the set of all the coupled models, the selection problem becomes 
finding the best coupled model in the set, $\mathcal{M}$, for prediction purposes.

\section{Bayesian Model Selection}
\label{sec:Bayes}


In the context of $\mathcal{M}$-closed perspective, the conventional Bayesian approach to 
model selection is to choose the model which has the highest posterior plausibility,
\begin{equation}
  M^* = \arg\max_{M} \pi(M|D,\mathcal{M}) .
\end{equation}
Given the data at the single physics level, one can compute the posterior model
plausibility for all the models in the $\mathcal{M}^A$ and $\mathcal{M}^B$ sets, as the
product between evidence and prior plausibility,
\begin{eqnarray}
  \pi(M_i^A|D^A,\mathcal{M}) \propto \pi(D^A|M_i^A,\mathcal{M})\pi(M_i^A|\mathcal{M}) .
\end{eqnarray}
The evidence is obtained during the calibration process for each single-physics models,
and it is given by the normalization constant in the Bayes rule, used to compute the 
posterior pdf of model parameters,
\begin{eqnarray}
  \pi({\param{\boldsymbol{\theta}}_i^A}|D^A,M_i^A,\mathcal{M}) = \frac{\pi(D^A|{\param{\boldsymbol{\theta}}_i^A},M_i^A,\mathcal{M}) \pi({\param{\boldsymbol{\theta}}_i^A}|M_i^A,\mathcal{M})}{\pi(D^A|M_i^A,\mathcal{M})} . \nonumber
\end{eqnarray}

With no data at the coupled level, one can easily obtained the posterior plausibility
for coupled models as,
\begin{eqnarray}
  \pi(M_{ij}|D,\mathcal{M}) = \pi(M_i^A|D^A,\mathcal{M}) \pi(M_j^B|D^B,\mathcal{M}) .
\end{eqnarray}
Notice, that the best coupled model is given by coupling the best models at the single physics level. Computationally this is advantageous as there is no need to build all the possible coupled models using the single physics models.
Looking at just two coupled models, we can say that we prefer model $M_1$ over
model $M_2$ if and only if $\pi(M_1|D,\mathcal{M}) > \pi(M_2|D,\mathcal{M})$. This inequality can be recasted as the following product of ratios:
\begin{equation}
\underbrace{\frac{\pi(D|M_1)}{\pi(D|M_2)}}_{\text{Bayes factor}} 
  \underbrace{\frac{\pi(M_1,\mathcal{M})}{\pi(M_2,\mathcal{M})}}_{\text{prior odds}} > 1
\end{equation}
The log-evidence is the trade-off between model complexity and
how well the model fits the data. In other words the evidence
yields the best model that obeys the law of parsimony,
\small
\begin{align}
  \ln [\pi(D|M_1)] =&
  E\left[\ln [\pi(D|{\param{\boldsymbol{\theta}}},M_1)]\right] \nonumber \\
&- \mathrm{KL}\bigg( \pi({\param{\boldsymbol{\theta}}}|D,M_1) ~||~ \pi({\param{\boldsymbol{\theta}}}|M_1) \bigg) ,
\end{align}    
\normalsize
where the model complexity is given by the Kullback-Leibler divergence
between posterior pdf and prior pdf of model parameters \cite{Cheung_2009C}. 
Therefore, this model selection scheme makes use of the following information in choosing
the best model: model complexity, data fit, and prior knowledge. Since it obeys Occam's 
razor, we implicitly gain some robustness with respect to predictions. However, if we have
two different QoIs that we would like to predict, it is not obvious if
the model selected under this scheme will be able to provide equally good predictions
for both QoIs. This is due to the fact that the information about the QoI is not 
explicitly used in the selection criterion. In the following sections, we will present
an extension of model selection scheme to also account for the QoI, and discuss its 
implications.

\section{Predictive Model Selection}
\label{sec:predictive}


Given a model class set $\mathcal{M}=\{M_1,M_2,\ldots,M_K\}$ (for simplicity the double index will be ignored in the model class notation), and a set of observations $D=\{\mathbf{d}_1,\mathbf{d}_2, \ldots, \mathbf{d}_n\}$ generated by an unknown model from $\mathcal{M}$, we are concerned with the problem of selecting the best model to predict an unobserved quantity of interest ${\qoi\mathbf{q}}$. 
The selection of the model which best estimates the pdf of the QoI is seen here as a decision problem~\cite{Trottini2002}. First one has to find the predictive distribution for each model class and the selection of the best model class is based on the utility of its predictive distribution. Given all the available information, the Bayesian predictive distribution conditioned on a model $M_j$ is given by:
\begin{equation}
\pi({\qoi\mathbf{q}}|D,M_{j})
=
\int \pi({\qoi\mathbf{q}}|{\param{\boldsymbol{\theta_j}}},D,M_{j})~
{\pi({\param{\boldsymbol{\theta_j}}}|D,M_{j})}~d{\param{\boldsymbol{\theta_j}}}
\end{equation}
where the posterior pdf for model parameters is computed using Bayes rule.
%
%
In the followings it is assumed that the true distribution of the QoI is generated by 
a model $M_j({\param\boldsymbol{\theta}_j}) \in M_j$, called the true model. Thus, the true pdf of the QoI can be 
written as:
\begin{equation}
\pi({\qoi\mathbf{q}} | {\param\boldsymbol{\theta}}, \mathbf{s}, D, \mathcal{M}) = \sum_{j=1}^K \pi({\qoi\mathbf{q}} | {\param\boldsymbol{\theta}_j}, D, M_j) s_j
\end{equation}
where ${\param\boldsymbol{\theta}} = ({\param\boldsymbol{\theta}_1},{\param\boldsymbol{\theta}_2},\ldots, {\param\boldsymbol{\theta}_K})$, $\mathbf{s} = (s_1,s_2,\ldots,s_K)$, and $s_j = 1$ if and only if the true model belongs to model class $M_j$.
Let $U_{\qoi\mathbf{q}}( {\param\boldsymbol{\theta}}, \mathbf{s},M_{j} )$ describe the 
utility of choosing the pdf associated with model class $M_j$ as the predictive 
distribution of the QoI, ${\qoi\mathbf{q}}$. Here the utility function is defined as 
the negative Kullback-Leibler divergence of the true distribution and the 
predictive distribution of $M_j$:
\small
\begin{align}
U_{\qoi\mathbf{q}}( {\param\boldsymbol{\theta}}, \mathbf{s},M_{j} ) =& - \mathrm{KL}\bigg( \pi({\qoi\mathbf{q}} | {\param\boldsymbol{\theta}}, \mathbf{s}, D, \mathcal{M}) ~||~ \pi({\qoi\mathbf{q}}|D,M_{j}) \bigg)
%
\end{align}
\normalsize

The model that maximizes the following expected utility (QoI-aware evidence)
is the model of choice for predictive purposes:
\begin{equation}
M^* = \arg\max_{M_j \in \mathcal{M}} \underbrace{\int U_{\qoi\mathbf{q}}( {\param\boldsymbol{\theta}}, \mathbf{s},M_{j} ) \pi({\param\boldsymbol{\theta}}, \mathbf{s} | D, \mathcal{M}) d{\param\boldsymbol{\theta}} d\mathbf{s}}_{E_{{\param\boldsymbol{\theta}}, \mathbf{s}}[U_{\qoi\mathbf{q}}( {\param\boldsymbol{\theta}}, \mathbf{s},M_{j} )]} \label{model_selection}
\end{equation}
With few mathematical manipulations, the expected utility can be written as follows,
\footnotesize
\begin{align}
&E_{{\param\boldsymbol{\theta}}, \mathbf{s}}[U_{\qoi\mathbf{q}}( {\param\boldsymbol{\theta}}, \mathbf{s},M_{j} )] = \int U_{\qoi\mathbf{q}}( {\param\boldsymbol{\theta}}, \mathbf{s},M_{j} ) \pi({\param\boldsymbol{\theta}} | \mathbf{s} , D, \mathcal{M}) \pi(\mathbf{s} | D, \mathcal{M}) d{\param\boldsymbol{\theta}} d\mathbf{s} \nonumber \\
&= \sum_{i=1}^K \pi(M_i|D,\mathcal{M}) \int U_{\qoi\mathbf{q}}( {\param\boldsymbol{\theta}_i}, s_i, M_{j} ) \pi({\param\boldsymbol{\theta}_i} | D, M_i) d{\param\boldsymbol{\theta}_i} \label{exp_util_sum} \\
&= -\sum_{i=1}^K \pi(M_i|D,\mathcal{M}) E_{{\param\boldsymbol{\theta}_i}}\bigg[ \mathrm{KL}\bigg( \pi({\qoi\mathbf{q}} | {\param\boldsymbol{\theta}_i}, D, M_i) || \pi({\qoi\mathbf{q}}|D,M_{j}) \bigg) \bigg] \nonumber
\end{align}
\normalsize
%

\subsection{Interpretation of the expected utility used in model selection}

Consider now that only two model classes exist in our model set $\mathcal{M} = \{ M_1,M_2 \}$. We prefer model class $M_1$ over $M_2$ and write $M_1 \succ M_2$ if and only if:
\begin{equation}
E_{{\param\boldsymbol{\theta}}, \mathbf{s}}[U_{\qoi\mathbf{q}}( {\param\boldsymbol{\theta}}, \mathbf{s},M_{1} )] > E_{{\param\boldsymbol{\theta}}, \mathbf{s}}[U_{\qoi\mathbf{q}}( {\param\boldsymbol{\theta}}, \mathbf{s},M_{2} )] \label{ineq}
\end{equation}

Substituting Eq.\eqref{exp_util_sum} into Eq.\eqref{ineq} the following model selection criterion can be derived:
\begin{eqnarray} \label{criterion}
\underbrace{\frac{R(M_1||M_2)}{R(M_2||M_1)}}_{\text{Risk ratio}} 
\underbrace{\frac{\pi(D|M_1)}{\pi(D|M_2)}}_{\text{Bayes factor}} 
\underbrace{\frac{\pi(M_1,\mathcal{M})}{\pi(M_2,\mathcal{M})}}_{\text{Prior odds}} > 1
\end{eqnarray}
where the numerator in the predictive risk ratio is given by the following expressions. The denominator is obtained by analogy with the numerator.
\small
\begin{eqnarray}
R(M_1||M_2) = E_{{\param\boldsymbol{\theta}_1}}\bigg[  \mathrm{KL}\bigg( \pi({\qoi\mathbf{q}} | {\param\boldsymbol{\theta}_1}, D, M_1) ~||~ \pi({\qoi\mathbf{q}}|D,M_{2}) \bigg) \bigg] \nonumber \\
- E_{{\param\boldsymbol{\theta}_1}}\bigg[  \mathrm{KL}\bigg( \pi({\qoi\mathbf{q}} | {\param\boldsymbol{\theta}_1}, D, M_1) ~||~ \pi({\qoi\mathbf{q}}|D,M_{1}) \bigg)  \bigg] 
%
%
\end{eqnarray}
\normalsize
%
%
%
The model selection criterion in Eq.\eqref{criterion} can be interpreted as the evidence of model class $M_1$ in favor of model class $M_2$, and is composed of prior evidence given by the prior odds, experimental evidence given by the Bayes factor and the predictive risk ratio which accounts for the loss of choosing the wrong model. According to Trottini and Spezzaferri~\cite{Trottini2002}, the expectations in the above ratio have the following meaning: {\small$E_{{\param\boldsymbol{\theta}_1}}\bigg[  \mathrm{KL}\bigg( \pi({\qoi\mathbf{q}} | {\param\boldsymbol{\theta}_1}, D, M_1) ~||~ \pi({\qoi\mathbf{q}}|D,M_{2}) \bigg) \bigg]$} - 
the risk of choosing model class $M_2$ when the true model belongs to $M_1$; {\small$E_{{\param\boldsymbol{\theta}_1}}\bigg[ \mathrm{KL}\bigg( \pi({\qoi\mathbf{q}} | {\param\boldsymbol{\theta}_1}, D, M_1) ~||~ \pi({\qoi\mathbf{q}}|D,M_{1}) \bigg) \bigg]$} - even if we report the distribution $\pi({\qoi\mathbf{q}}|D,M_{1})$ when the true model belongs to $M_1$, there is a risk incurred due to the unknown value of ${\param\boldsymbol{\theta}_1}$ that generated the true model. Comparing with the previous model selection scheme, the following information is used in this scheme to select the best model: QoI, model complexity, data fit, and prior knowledge.

%




\subsection{Calculating the expected utility used in model selection}

The calculation of the QoI-aware evidence in Eq.\eqref{exp_util_sum} is challenging as we are dealing with high dimensional integrals, and the number of samples in the posterior distributions is dependent on the MCMC algorithms and computational complexity of the forward model. Thus, we would like to simplify this calculation. Starting from Eq.\eqref{exp_util_sum} the following expression for the expected utility
can be obtained:

\begin{strip}
\rule[0in]{3.3in}{.001in}
\small
\begin{align}
E_{{\param\boldsymbol{\theta}}, \mathbf{s}}[U_{\qoi\mathbf{q}}( {\param\boldsymbol{\theta}}, \mathbf{s},M_{j} )]
%
%
&= - \sum_{i=1}^K \pi(M_i|D,\mathcal{M}) \int \pi({\param\boldsymbol{\theta}_i} | D, M_i) 
\pi({\qoi\mathbf{q}} | {\param\boldsymbol{\theta}_i}, D, M_i) \log 
\frac{ \pi({\qoi\mathbf{q}} | {\param\boldsymbol{\theta}_i}, D, M_i) }
{ \pi({\qoi\mathbf{q}}|D,M_{j}) }
d{\param\boldsymbol{\theta}_i} d{{\qoi\mathbf{q}}} \nonumber \\
%
%
%
&= - \sum_{i=1}^K \pi(M_i|D,\mathcal{M}) \int \pi({\qoi\mathbf{q}},{\param\boldsymbol{\theta}_i} | D, M_i) 
\log \pi({\qoi\mathbf{q}} | {\param\boldsymbol{\theta}_i}, D, M_i)
d{\param\boldsymbol{\theta}_i} d{{\qoi\mathbf{q}}} +
 \int \pi({\qoi\mathbf{q}} | D, \mathcal{M}) \log \pi({\qoi\mathbf{q}}|D,M_{j}) d{\qoi\mathbf{q}}  \label{kl_exp_util}
\end{align} 
\normalsize
\begin{flushright}
\rule[0in]{3.3in}{.001in}
\end{flushright}
\end{strip}

Where the predictive pdf under all models is given by,
\begin{equation}
\pi({\qoi\mathbf{q}} | D, \mathcal{M}) = \sum_{i=1}^K \pi(M_i|D,\mathcal{M}) \pi({\qoi\mathbf{q}}|D,M_{i}) .
\end{equation}
Since the first term in Eq.\eqref{kl_exp_util} is the same for all models $M_j$, for $j = 1 \ldots K$, 
the optimization in Eq.\eqref{model_selection} is equivalent with maximizing the second term in Eq.\eqref{kl_exp_util}, which is the negative cross-entropy between
the predictive distribution conditioned on all the models and the predictive distribution conditioned 
on the $j$th model:
\begin{align}
M^* = \arg\max_{M_j \in \mathcal{M}} -\mathrm{H} \bigg( \pi({\qoi\mathbf{q}} | D, \mathcal{M}) , \pi({\qoi\mathbf{q}}|D,M_{j}) \bigg) \label{model_selection_interim}
\end{align}
By writing the optimization as a minimization instead of a maximization and subtracting the entropy
of the predictive distribution conditioned on all the models, then one can rewrite the model selection
problem as,
\footnotesize
\begin{align}
M^* &= \arg\min_{M_j \in \mathcal{M}} \mathrm{H} \bigg( \pi({\qoi\mathbf{q}} | D, \mathcal{M}) , \pi({\qoi\mathbf{q}}|D,M_{j}) \bigg) - \mathrm{H} \bigg( \pi({\qoi\mathbf{q}} | D, \mathcal{M}) \bigg) \nonumber \\
&= \arg\min_{M_j \in \mathcal{M}} \mathrm{KL}\bigg( \pi({\qoi\mathbf{q}} | D, \mathcal{M}) \bigg|\bigg| \pi({\qoi\mathbf{q}}|D,M_{j}) \bigg) 
\label{model_selection_final}
\end{align}
\normalsize

Thus, the best model to predict an unobserved quantity of interest ${\qoi\mathbf{q}}$ 
is the one whose predictive distribution best approximates the predictive distribution 
conditioned on all the models. This is rather intuitive as all we can say about the 
unobserved quantity of interest is encoded in the predictive distribution 
conditioned on all the models. The predictive pdf under all models being the most robust
estimate of the QoI for this problem. 

This model selection scheme reveals that when the posterior model plausibility is not able to discriminate between the models, the prediction obtained with the selected model is the most robust prediction we can obtain with one model. Thus we are able to account for model uncertainty when predicting the QoI. On the other hand, in the limit, for discriminatory observations, when the posterior plausibility is one for one of the models, the two selection schemes become equivalent. 

\section{Model Problem}
\label{sec:model}

The model problem consists of a spring-mass-damper system that is driven by an 
external force. The spring-mass-damper and the forcing function are considered to be
separate physics such that the full system model consists of a coupling of the 
dynamical system modeling the spring-mass-damper system and a function modeling the 
forcing. In this model problem, synthetic data are generated according to a truth system.

\subsection{Models}

This section describes the models that will form the sets of interest for the single
physics.
%
%
The models of the spring-mass-damper system take the following form:
\begin{equation}
m \ddot{x} + {\param c} \dot{x} + \tilde{k}(x) x = 0.
\label{eqn:smd_model}
\end{equation}
The mass is assumed to be perfectly known, $m=1$, and the damping
coefficient ${\param c}$ is a calibration parameter.  Model form uncertainty is
introduced through the spring models $\tilde{k}(x)$.  Three models are
considered: a linear spring ({\bf OLS}), a cubic spring ({\bf OCS}),
 and a quintic spring ({\bf OQS}), given by the following relations:
\begin{eqnarray}
\tilde{k}_{OLS}(x) &=& {\param k_{1,0}} \label{eqn:linear_spring} \\
\tilde{k}_{OCS}(x) &=& {\param k_{3,0}} + {\param k_{3,2}} x^2 \label{eqn:cubic_spring} \\
\tilde{k}_{OQS}(x) &=& {\param k_{5,0}} + {\param k_{5,2}} x^2 + {\param k_{5,4}} x^4 \label{eqn:quintic_spring}
\end{eqnarray}


The models of the forcing function are denoted $\tilde{f}(t)$.  Three 
models are considered: simple exponential decay ({\bf SED}), 
oscillatory linear decay ({\bf OLD}), and oscillatory exponential decay ({\bf OED}):
\small
\begin{align}
\tilde{f}_{SED}(t) &= {\param F_0} \exp(-t/{\param \tau}) \\
\tilde{f}_{OLD}(t) &=
\left\{ \begin{array}{c} 
{\param F_0} (1 - t/{\param \tau}) \left[ {\param \alpha} \sin({\param \omega} t) + 1 \right], \,\, 0 \leq t \leq {\param \tau} \\
0, \quad t > {\param \tau}
\end{array} \right. \\
\tilde{f}_{OED}(t) &= {\param F_0} \exp(-t/{\param \tau}) \left[ {\param \alpha} \sin({\param \omega} t) + 1 \right] \label{eqn:truth_forcing}
\end{align}
\normalsize


The coupling of the spring-mass-damper and the forcing is trivial.
Thus, only a single coupling model is considered, and the coupled
model is given by
\begin{equation}
m \ddot{x} + {\param c} \dot{x} + \tilde{k}(x) x = \tilde{f}(t).
\end{equation}
There are three choices for $\tilde{k}(x)$ and three choices for
$\tilde{f}(t)$, leading to nine total coupled models.

\subsection{The Truth System}

To evaluate the two selection schemes: Bayesian model selection, and
predictive model selection, we can construct the true system which will be 
used to generate data and give the true value of the QoI. The comparison
of the two selection criteria will be done with respect to different
subsets of models, and the ability of the best model to predict the true 
value of the QoI. The true model is described by {\bf OQS-OED}:
\begin{equation}
m \ddot{x} + c \dot{x} + \tilde{k}_{OQS}(x) x = \tilde{f}_{OED} \label{eqn:truth_smd_ode}
\end{equation}
where $m = 1$, $c = 0.1$. The parameters for the spring model are set to $k_{5,0} = 4$, 
$k_{5,2} = -5$, and $k_{5,4} = 1$. The true forcing function is given by the following
values for the parameters: $F_0 = 1$, $\tau = 2 \pi$, $\alpha = 0.2$, $\omega = 2$.


The QoI of the coupled model is assumed to be the maximum velocity 
$\dot{x}_{max} = \max_{t \in \mbb{R}^+} |\dot{x}(t)|$.
%
%
The observable for physics A is given by the kinetic energy versus time: $\frac{1}{2} \dot{x}(t_i)^2$ for $i=1, \ldots, N$.  Note that this contains the same information as the velocity except that it is ambiguous with respect to the sign. The observable for physics B is given by the force versus time: $f(t_i)$ for $i = 1, \ldots, M$. In both cases simulated observations have been generated by perturbing the deterministic predictions of the true model, with a log-normal multiplicative noise with standard deviation of $0.1$.

\section{Numerical Results}
\label{sec:results}

The inverse problem of calibrating the model parameters from the measurement data is solved
using MCMC simulations. In our simulations, samples from the posterior 
distribution are obtained using the
statistical library QUESO \cite{PrSc11, Cheung_2009C} equipped with the Hybrid Gibbs
Transitional Markov Chain Monte Carlo method proposed in Ref. \cite{Cheung_2008B}. One 
advantage of this MCMC algorithm is that it provides an accurate estimate of
the log-evidence using the adaptive thermodynamic integration.
Estimators based on $k$-nearest neighbor are used to compute the Kullback-Leibler divergence in Eq.\eqref{model_selection_final}, see Appendix. The use of these estimators is advantageous especially when only samples are available to describe the underlying distributions.

Three different scenarios are constructed to assess the predictive capability of the models selected using the two selections schemes: Bayesian model selection and predictive model selection. All the uncertain parameters of the models are considered uniformly distributed and the model error has also been calibrated and propagated to the QoI.

\begin{figure}[hp]
\centering
\subfigure[Case $1$: QoI pdf]{\includegraphics[height=.44\linewidth]{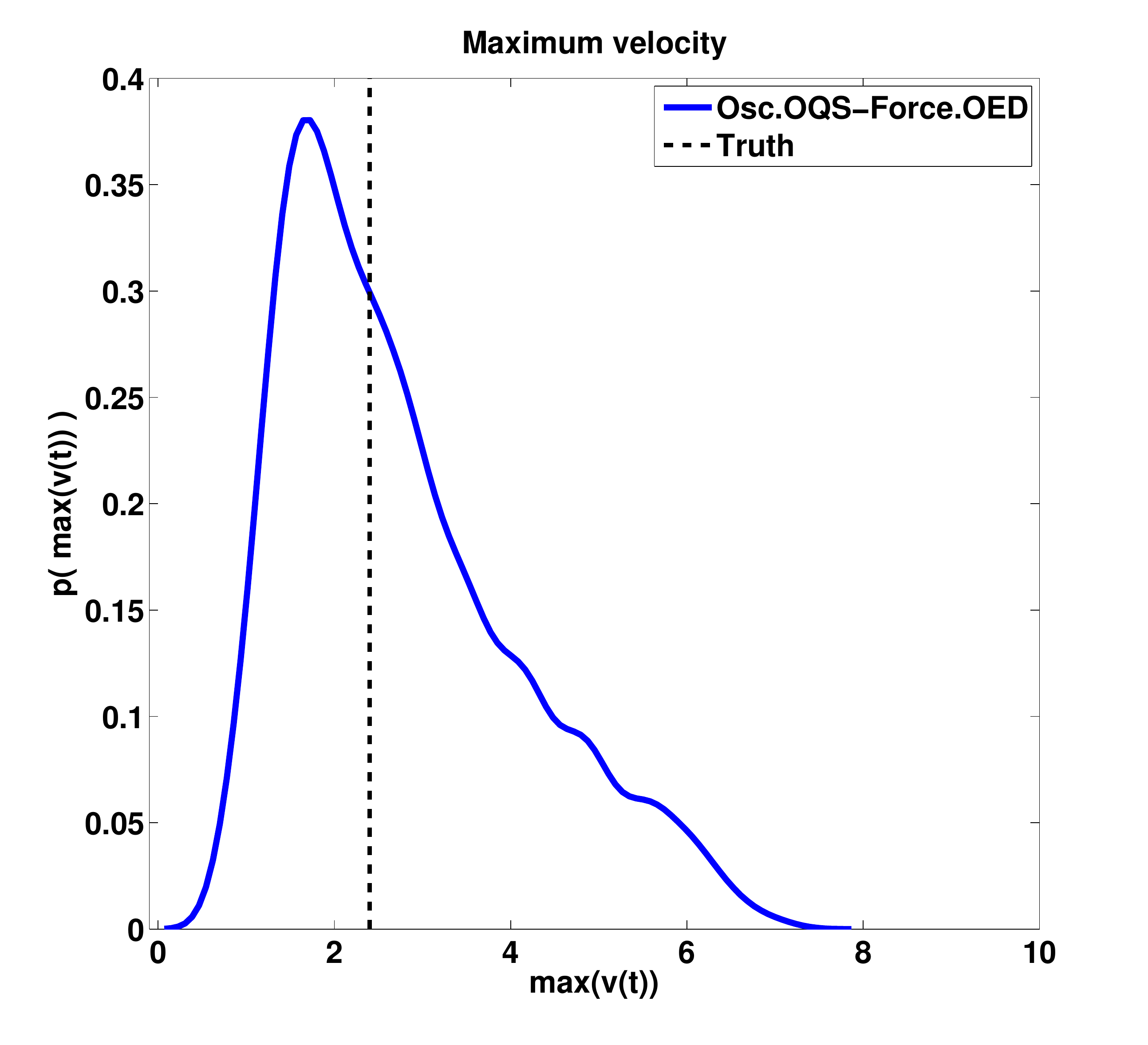}}
\subfigure[Case $2$: QoI pdfs]{\includegraphics[height=.44\linewidth]{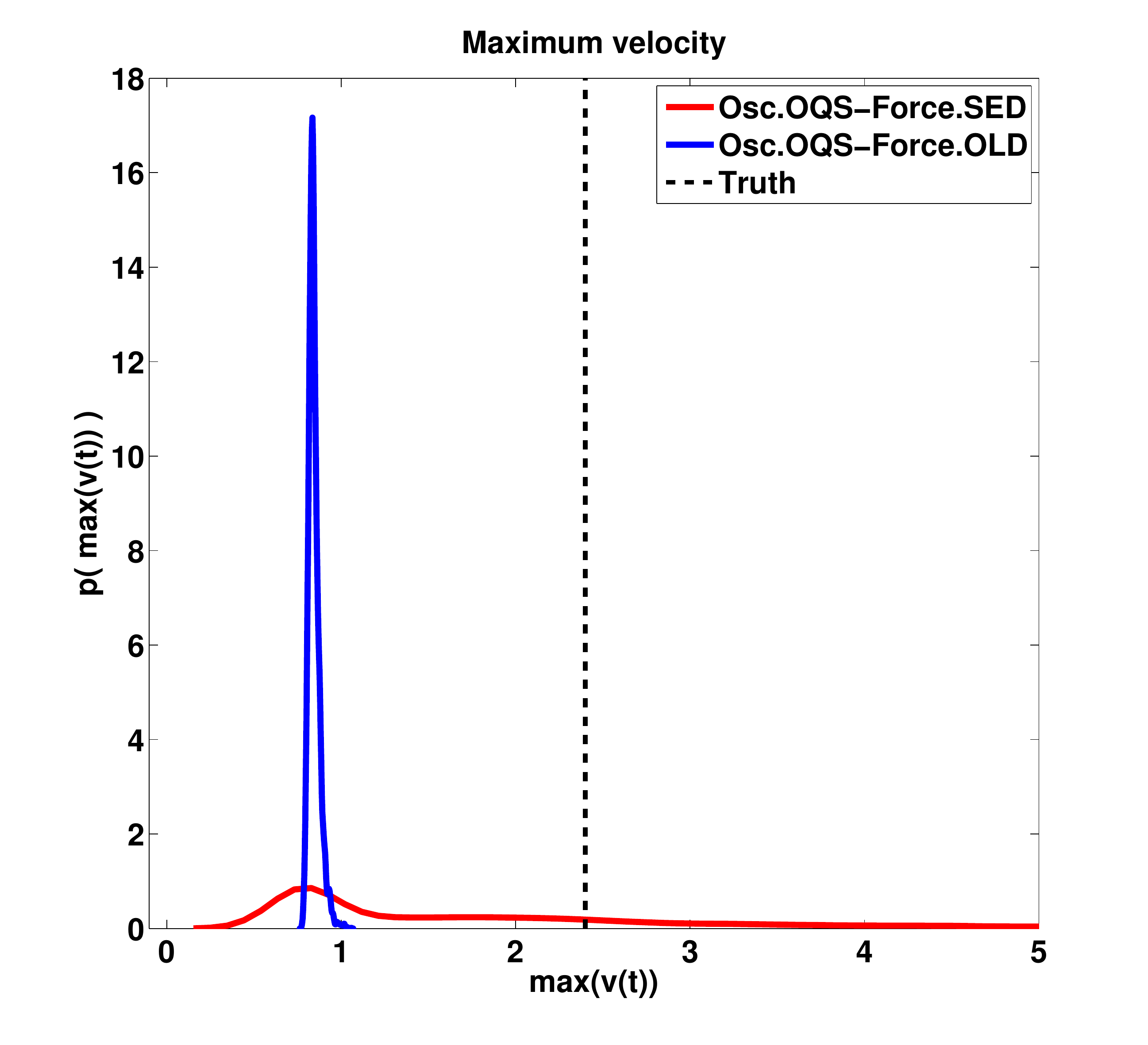}}
\subfigure[Case $2$: Observable]{\includegraphics[height=.44\linewidth]{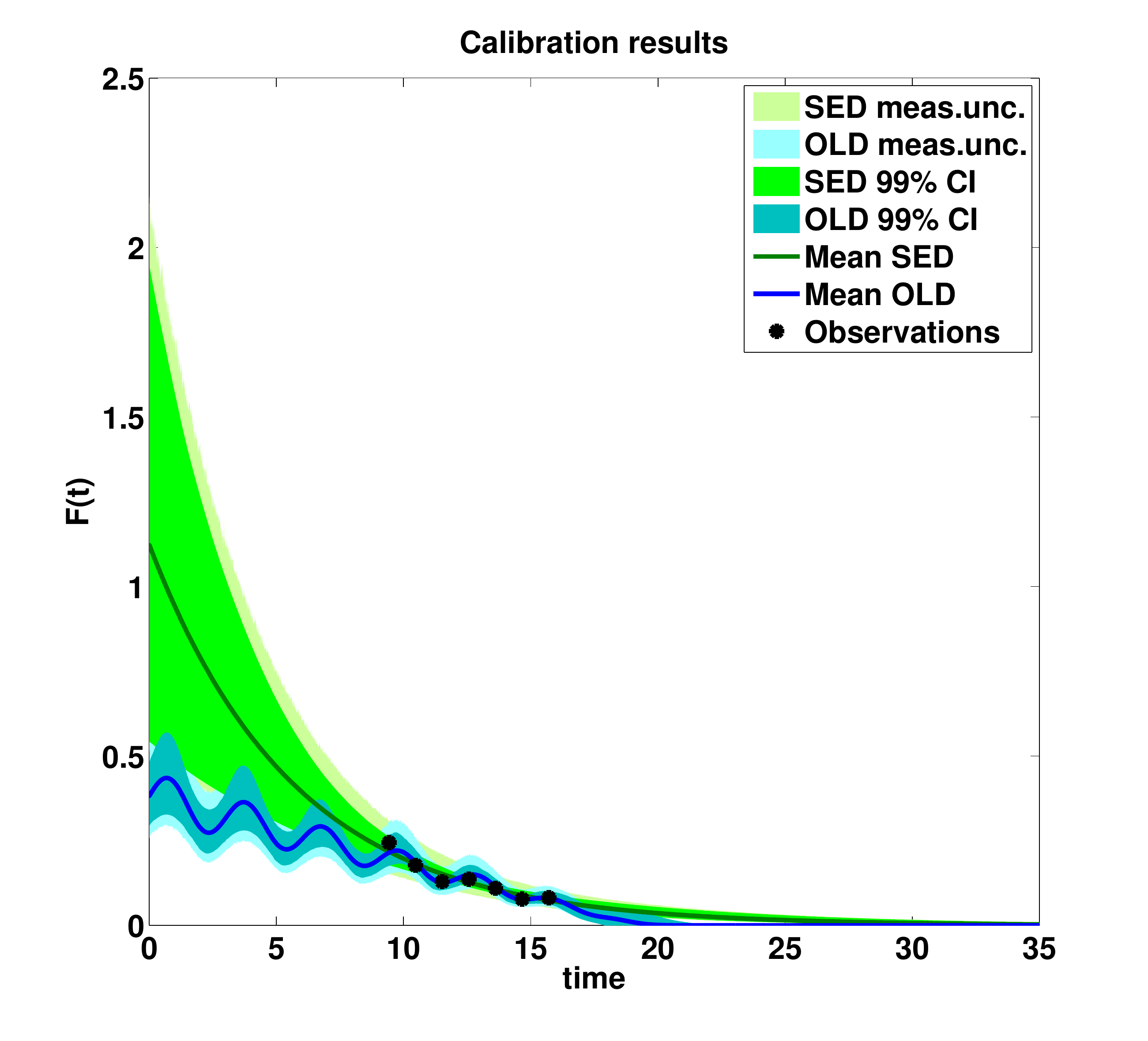}}
\subfigure[Case $3$: QoI pdfs]{\includegraphics[height=.44\linewidth]{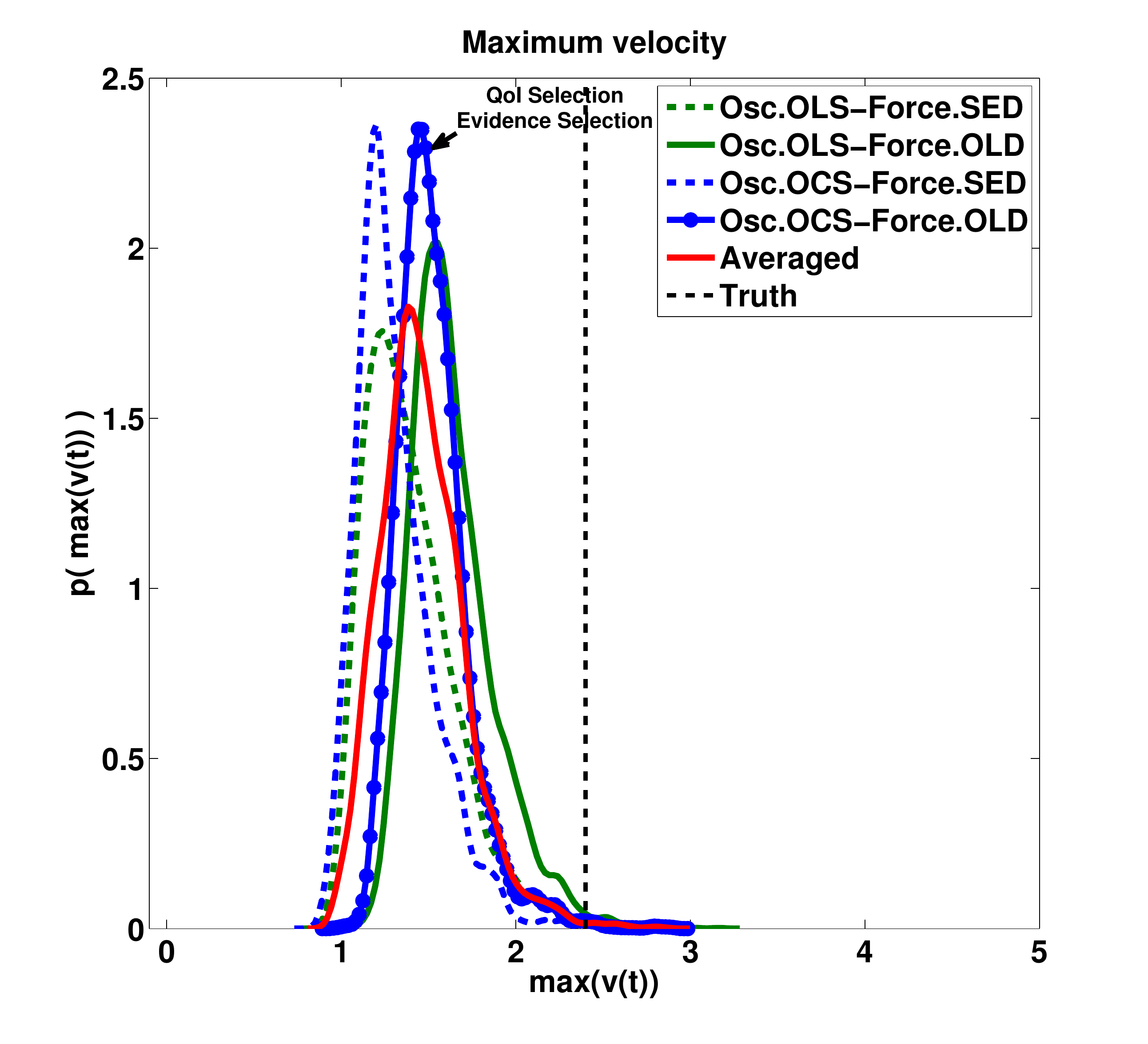}}
\caption{Results for the three cases considered}\label{fig:all}
\end{figure}

First, all the models are included in the two sets, including the components used to generate the true model. For oscillators the model class set is given by $\mathcal{M}^A = \{ M_{OLS}, M_{OCS}, M_{OQS} \}$ and for forcing function $\mathcal{M}^B = \{ M_{SED}, M_{OLD}, M_{OED} \}$. A number of $10$ measurements have been considered for the oscillators and $61$ for the forcing. Table \ref{tbl:ex1} summarizes the results obtained after applying the two approaches. On the first column and the first row, under each model one can find the model plausibility after calibration, and the first number in a cell gives the plausibility of the corresponding coupled model. The number in the parenthesis is the KL divergence used in the predictive model selection. In this case, the observations provided are enough to discriminate the models at single physics level, oscillator {\bf OQS} and the forcing {\bf OED} are selected in this case. Here, the predictive model selection is consistent with the plausibility based model selection. Notice that the true model belongs to the model class {\bf OQS-OED}, and the prediction of the selected model covers the true value of the QoI, see Fig.\ref{fig:all}a. One computational advantage is that when discriminatory observations are available, one does not need to carry the analysis on all the coupled models, just on the coupling of the best single physics ones and still be in agreement with predictive requirements. We argue that for very complex and hierarchical systems, with multiple levels of coupling, such a situation should be preferred and exploited by designing experiments to collect measurements intended to discriminate models.

\begin{table}[htbp]
  \centering
  \caption{Results Case $1$}
  \label{tbl:ex1}
  \begin{tabular}{|c|c|c|c|}
    \hline
    ~ & {\bf SED} & {\bf OLD} & {\bf OED*} \\
    ~ & $0.00$ & $0.00$ & {\color{blue}$1.00$} \\
    \hline
    {\bf OLS} &  $0.00$ & $0.00$ & $0.00$ \\
    $0.00$ &  $(3.75)$ & $(3.36)$ & $(3.67)$ \\
    \hline
    {\bf OCS} &  $0.00$ & $0.00$ & $0.00$ \\
    $0.00$ &  $(4.19)$ & $(3.91)$ & $(4.24)$ \\
    \hline
    {\bf OQS*} &  $0.00$ & $0.00$ & {\color{blue}$1.00$} \\
    {\color{blue}$1.00$} &  $(1.15)$ & $(5.09)$ & {\color{blue}$(0.00)$} \\
    \hline
  \end{tabular}
\end{table}
For the second scenario we remove the forcing that generated the true model. Now the sets of model classes are given by $\mathcal{M}^A = \{ M_{OLS}, M_{OCS}, M_{OQS} \}$ and $\mathcal{M}^B = \{ M_{SED}, M_{OLD} \}$. The same number of observations are considered for the oscillators and $7$ measurements for the forcing models. The results are presented in Table \ref{tbl:ex2}. As before, we are able to discriminate the oscillators, however we cannot say the same for the forcing models. This can be seen in Fig. \ref{fig:all}c, where we can see the prediction of the observable with the two forcing models after calibration. The data supports almost equally well both forcing functions. In this case the two selection schemes yield two different models: {\bf OQS-OLD} for the Bayesian model selection and {\bf OQS-SED} for the predictive model selection, see Table \ref{tbl:ex2}. Looking at their predictions for the QoI, Fig. \ref{fig:all}b, we see that the pdf provided by the model selected using conventional Bayesian model selection doesn't even cover the true value of the QoI, whereas the one chosen by the predictive selection scheme covers the true value of the QoI in the tail. Thus, while the plausibility based selection ignores the model uncertainty, the predictive selection approach yields the model with the most robust predictive pdf for the QoI. This prediction incorporates as much as possible model uncertainty that one can obtain with just one model.
Therefore, the predictive approach is recommended in the case when discriminatory observations are not available and one model has to be chosen instead of model averaging, especially for complex hierarchical systems.
%
%
\begin{table}[htbp]
\begin{minipage}[b]{0.45\linewidth}
  \centering
  \caption{Results Case $2$}
  \label{tbl:ex2}
  \begin{tabular}{|c|c|c|}
    \hline
    ~ & {\bf SED} & {\bf OLD} \\
    ~ & $0.44$ & {\color{blue}$0.56$} \\
    \hline
    {\bf OLS} &  $0.00$ & $0.00$ \\
    $0.00$ &  $(6.77)$ & $(7.35)$ \\
    \hline
    {\bf OCS} &  $0.00$ & $0.00$ \\
    $0.00$ &  $(5.66)$ & $(7.23)$ \\
    \hline
    {\bf OQS*} &  $0.44$ & {\color{blue}$0.56$} \\
    {\color{blue}$1.00$} & {\color{red}$(0.62)$} & $(1.64)$ \\
    \hline
  \end{tabular}
\end{minipage}
\hspace{0.5cm}
\begin{minipage}[b]{0.45\linewidth}
  \centering
  \caption{Results Case $3$}
  \label{tbl:ex3}
  \begin{tabular}{|c|c|c|}
    \hline
    ~ & {\bf SED} & {\bf OLD} \\
    ~ & $0.48$ & {\color{blue}$0.52$} \\
    \hline
    {\bf OLS} &  $0.05$ & $0.06$ \\
    $0.12$ &  $(0.39)$ & $(0.82)$ \\
    \hline
    {\bf OCS} &  $0.42$ & {\color{blue}$0.45$} \\
    {\color{blue}$0.88$} &  $(0.37)$ & {\color{blue}$(0.34)$} \\
    \hline
  \end{tabular}
\end{minipage}
\end{table}


Lastly, we are not including in the model sets any of the components that generated
the true model. Thus, $\mathcal{M}^A = \{ M_{OLS}, M_{OCS} \}$ and $\mathcal{M}^B = \{ M_{SED}, M_{OLD} \}$. In this case only $5$ observations are considered for the oscillators and $4$ for the forcing functions. We can see from Table \ref{tbl:ex3} that at the single physics level we are not able to discriminate the oscillators or the forcing functions. For the coupled models both approaches choose the same model {\bf OCS-OLD}, however looking at the predictions of all coupled models, including their average prediction we see that all of them give very low likelihood for the true value of the QoI. This case emphasizes that the prediction approach has the same drawback as the Bayesian model selection and Bayesian model averaging. Mainly, we are at the mercy of our hypotheses and the only way to escape this case is to generate additional hypotheses.

%

\section{Conclusions}
\label{sec:conclusions}

In this paper, we have presented a model selection criterion that accounts for
the predictive capability of coupled models. This is especially useful when 
complex/multiphysics models are used to calculate a quantity of interest. 
It has been shown that the prediction obtain with the model chosen by the 
predictive approach, is the most robust prediction that one can obtain with 
just one model. This is because in part it incorporates model uncertainty,
while conventional Bayesian model selection ignores it. 
For discriminatory measurements the Bayesian model
selection and predictive model selection are equivalent. This suggests that when
additional data collection is possible then designing experiments for model
discrimination is computationally preferred for complex models.

\section*{Appendix}

The approximation of the Kullback-Leibler 
divergence is based on a $k$-nearest neighbor approach~\cite{Wang2006}. 
\small
\begin{eqnarray}
  \mathrm{KL}\bigg( p(x|D_n) \bigg|\bigg| p(x|D_{n-1}) \bigg) \approx \frac{d_X}{N_n} \sum_{i=1}^{N_n} 
  \log \frac{\nu_{n-1}(i)}{\rho_n(i)} + \log \frac{N_{n-1}}{N_n-1} \nonumber
\end{eqnarray}
\normalsize
where $d_X$ is the dimensionality of the random variable $X$, $N_n$ and $N_{n-1}$ give the 
number of samples $\{X_n^i|i=1,\ldots,N_n\} \sim p(x|D_n)$ and 
$\{X_{n-1}^j|j=1,\ldots,N_{n-1}\} \sim p(x|D_{n-1})$ respectively, and the two distances 
$\nu_{n-1}(i)$ and $\rho_n(i)$ are defined as follows:
$\rho_n(i) = \min_{j=1\ldots N_n, j \ne i} \| X_n^i - X_n^j \|_\infty$ and 
$\nu_{n-1}(i) = \min _{j=1\ldots N_{n-1}} \| X_n^i - X_{n-1}^j \|_\infty$.

\section*{Acknowledgments}

This material is based upon work supported by the Department of 
Energy [National Nuclear Security Administration] under Award Number [DE-FC52-08NA28615].


\bibliographystyle{plain}
\bibliography{main}


\end{document}